\documentclass[manuscript]{aastex62}
\usepackage{graphicx}
\usepackage{epstopdf}

\shorttitle{On the Interpretation of XSPEC Abundances and Emission Measures}
\shortauthors{Leahy, Foster \& Seitenzahl}

\begin{document}


\title{On the Interpretation of XSPEC Abundances and Emission Measures}


\author{D.A. Leahy}
\affil{Department of Physics $\&$ Astronomy, University of Calgary, Calgary,
Alberta T2N 1N4, Canada}
\author{A. Foster}
\affil{Center for Astrophysics, Harvard \& Smithsonian
60 Garden Street MS-4, Cambridge, MA 02138, USA
}
\author{I. Seitenzahl}
\affil{UNSW Canberra, School of Science
Canberra, Australia}



\begin{abstract}
{The purpose of this work is to describe the assumptions built into the X-ray spectrum fitting software XSPEC for the calculation of element abundances and emission measure of a plasma and to describe the effects when those assumptions are not accurate. 
The ratio of electron density to hydrogen density in XSPEC is fixed at a constant.
The correct ratio can be calculated from the ionization states of the elements. 
We show the constant value used in XSPEC is valid to within $\simeq3.5$\% for a solar abundance plasma.
For a plasma that deviates from solar abundance, such as hydrogen-poor or heavy element rich plasmas as found in the ejecta of supernova remnants, this ratio can smaller by factors of 0.1 to 0.001.
The hydrogen emission measure, defined by integral of electron density times hydrogen density over plasma volume, is derived from the $norm$ in XSPEC, but one needs to include the hydrogen abundance factor. 
For other elements, the emission measures are the XSPEC values multiplied by the element abundance factors.
Using the correct electron-to-hydrogen ratio and emission measures, we show the correct electron density is smaller by the square root of the correct electron density ratio divided by the XSPEC value. 
Element densities and total masses (for given distance and volume) are larger by the abundance factors divided by the above square root. 
Because hydrogen-poor plasmas occur in the ejecta of Type Ia supernova remnants, previously estimated element masses from X-ray spectra are likely significantly underestimated.
}
\end{abstract}




\section{Introduction}

Spectral model fits to observed spectra of hot (X-ray emitting) plasmas provide us crucial information
on the state and amount of hot plasma.
The spectral fits inform us of the elemental composition of the plasma and its temperature and, in the case of non-equlibrium ionization plasmas, tell us the age of the plasma by the parameter $\tau=n_e~t$, with $t$ the age since the plasma was shock heated.

An astrophysical example of hot plasmas of great interest are the shock-heated plasmas in a supernova remnant.
For young supernova remnants there is significant emission from the forward shock into the circumstellar or interstellar medium and from the reverse shock into the ejected material of the progenitor star. 
Thus, for example, we learn about the supernova progenitor and the explosion mechanism by studying the composition of the ejecta via fitting hot plasma emission models to the observed X-ray spectrum of the shock-heated ejecta.


The most common modeling software for X-ray spectra is the XSPEC software package \citep{1996ASPC..101...17A}\footnote{Available at HEASARC at https://heasarc.gsfc.nasa.gov/docs/xanadu/xanadu.html}. 
XSPEC has built-in assumptions which affect the interpretation of emission measures. 
The purpose of the current work is to explain the assumptions and provide corrections for cases where the assumptions are not accurate.
This issue was previously considered by \cite{2022Univ....8..274L}. 
That work developed an approximate correction to emission measures from XSPEC based on calculating mean molecular weights.
Here we develop the more accurate exact expressions based on the detailed assumptions used by XSPEC in calculating emission measures. 

Our analysis 
starts with a brief overview of the XSPEC hot plasma spectrum calculation Section~\ref{sec:xspec}.
Then we make general remarks and give definitions for emission measure in Section~\ref{sec:EMgen}.
Section~\ref{sec:zdens} gives how XSPEC defines element and electron densities. 
Calculation of electron density for solar abundance cases is given in Section~\ref{sec:nesol},
and for non-solar abundance in Section~\ref{sec:nenonsol}.
Section~\ref{sec:nion} explains how ion densities are derived from XSPEC $norm$ using the fundamental formula for emissivity.
We distinguish between constrained and unconstrained elements in Section~\ref{sec:spec}, where unconstrained elements are those with abundances not constrained by the spectral fit. 
In Section~\ref{sec:spec}, we present simulated spectral fits to illustrate typical upper limits for abundances of H and He for hydrogen poor plasmas.
Section~\ref{sec:results} presents the results of this work. This includes presentation of the general formulae (valid for solar and non-solar abundances) for emission measure (Section~\ref{sec:EM}), 
for element densities (Section~\ref{sec:nZ}),
and for element masses (Section~\ref{sec:Zmass}).
The above formulae include the effect of using the correct electron to hydrogen density ratio and element abundance factors. 
The conclusions are summarized in Section~\ref{sec:sum}.

\section{Analysis}\label{sec:analysis}

\subsection{XSPEC hot plasma calculation}\label{sec:xspec}

Here we consider hot plasma emission models in XSPEC. 
These include $nei$, $vnei$, and $vvnei$ models, where $nei$ has solar abundances, $vnei$ has abundances of He, C, N, O, Ne, Mg, Si, S, Ar, Ca, Fe, and Ni specified by 12 parameters and $vvnei$ has abundances of elements 1$\le Z \le$30  specified by 30 parameters. 
This discussion applies to other hot plasma emission models ($apec$, $bapec$, $bremms$, $meka$, $mekal$, and  $raymond$) and their $v$ and $vv$ generalizations.

\subsection{Emission measure, general remarks}\label{sec:EMgen}

The primary standard definition of emission measure $EM$ uses electron density, $n_e$:
\begin{equation}
 EM_e=\int n_e(\vec{r})^2 dV =n_e^2 ~V~~\rm{(if~uniform)}  \label{eqnEMe}
 \end{equation}
where  $\vec{r}$ is the 3-dimensional position vector with arbitrary origin (e.g. at the observer or at center of emitting volume) and the integration is over the volume, $V$, of the emitting gas. 
In the last expression,  a uniform density plasma is assumed for simplicity. 
Another definition of emission measure $EM$ is:
\begin{equation}
 EM_H=\int n_e(\vec{r}) n_{H}(\vec{r}) dV =n_e ~ n_{H} ~V~~\rm{(if~uniform)}  \label{eqnEM}
 \end{equation}
with $n_{H}$ the number density hydrogen nuclei.
It is straightforward to generalize to variable densities using $n_e(\vec{r})=n_e~ g_e(\vec{r})$ and $n_H(\vec{r})=n_H~ g_H(\vec{r})$, 
with the normalization\footnote{If  $g_e(\vec{r})=g_H(\vec{r})$ this fully specifies the normalizations, otherwise one can normalize one function e.g. by $\int g_e(\vec{r})^2 dV=V$, so the two normalization conditions fix both function normalizations.} of $g_e(\vec{r})$ and $g_H(\vec{r})$ chosen such that
$\int g_e(\vec{r})~g_H(\vec{r}) dV = V$.

Because the second definition above is not useful when the hydrogen abundance is zero, we introduce $EM_Z$ for each element $Z$:
\begin{equation}
 EM_Z=\int n_e(\vec{r}) n_{Z}(\vec{r}) dV= n_e~ n_Z~ V ~~\rm{(if~uniform)}  \label{eqnEMZ}
 \end{equation}
 Then $EM_Z$ can be non-zero for $Z>1$ even if there is no hydrogen.
 There is an $EM_N$ based on the total number of nuclei $n_N=\sum_{Z=1}^{Z_{max}} n_Z$: 
\begin{equation}
 EM_N=\int n_e(\vec{r}) n_{N}(\vec{r}) dV= n_e~ n_N~ V ~~\rm{(if~uniform)}  \label{eqnEMN}
 \end{equation}
 $EM_N$ is always non-zero.
 The other types of $EM$ can be found from $EM_N$ using the above definitions\footnote{This also holds for the nonuniform case because of the choice of normalizations of the $g(\vec{r})$ functions:
  $EM_H=\frac{n_H}{n_N}~EM_N~(\int g_e(\vec{r})~g_H(\vec{r}) dV)/(\int g_e(\vec{r})~g_N(\vec{r}) dV)=\frac{n_H}{n_N}~EM_N$.},
 e.g.  $EM_H=\frac{n_H}{n_N}~EM_N$ or  $EM_e=\frac{n_e}{n_N}~EM_N$ 

For XSPEC, the value of the electron density, $n_{e,X}$,  is tied to the hydrogen density, $n_{H,X}$,  by: 
 \begin{equation}
n_{e,X}=1.2~n_{H,X}  \label{eqnne}
\end{equation}
The subscript $X$ indicates XSPEC values.
The above value (1.2) is the ratio of electron density to hydrogen density for a solar abundance plasma (this is discussed in more detail in Section~\ref{sec:nesol} below).
Generally, the  electron to hydrogen ratio depends on the charge states of all the elements, which depend on the plasma conditions of composition, temperature and amount of non-equilibrium ionization.
However it doesn't vary much for a solar abundance plasma which is dominated by hydrogen and helium.
The adoption of a fixed ratio makes the fitting process much faster because it avoids having to recalculate the electron to hydrogen density ratio every time the plasma conditions are changed.

The XSPEC norm for hot plasma emission models is proportional to $EM_{H,X}$: 
\begin{eqnarray}
 norm_X&=&10^{-14}~EM_{H,X} /(4\pi [D_A(1+z)]^2) ~~~\rm{with} \nonumber \\
 EM_{H,X}&=&\int n_{e,X}(\vec{r})~n_{H,X}(\vec{r})~ dV =n_{e,X}~n_{H,X}~V~~\rm{(if~uniform)}
 \label{eqnXnorm}
\end{eqnarray}
with $EM_{H,X}$ in units of cm$^{-3}$, 
$z$ is redshift and $D_A$ is angular diameter distance  in units of cm. 
A derivation of the relation between $norm_X$ and $EM_{X<H}$ is given in Section~\ref{sec:nion}.
In the next section we discuss the relations between $n_{H,X}$ and $n_H$, 
between $n_{e,X}$ and $n_e$, and how they relate to element densities.

\subsection{Element and electron densities in XSPEC}\label{sec:zdens}

In XSPEC there are nine choices available for solar abundances, defined by $A_{sol,Z}$=$n_{Z,sol}/ n_{H}$ with $A_{sol,Z}$ is the number ratio of element $Z$ to hydrogen for solar abundance. 
The abund command in XSPEC determines which set of solar abundance is used, with the default set as `angr' (see the XSPEC manual at  https://heasarc.gsfc.nasa.gov/docs/xanadu/xspec/manual/manual.html for details).

Abundances for elements $Z$ in XSPEC are specified by abundance factors $A_{Z}$ which are relative to solar abundance. 
However to allow for independent adjustment in the hydrogen abundance in XSPEC, a fiducial hydrogen density $n_{H,0}$ is defined, with:
 \begin{equation}
n_{H}= A_1~n_{H,0} \label{eqnnH}
 \end{equation} 
Other elements abundances are also defined in terms of $n_{H,0}$:
\begin{equation}
n_{Z}= A_{sol,Z} ~ A_{Z}~n_{H,0} \label{eqnnZ}
 \end{equation} 
 Thus $A_{Z}$=1 for all $Z$ for solar abundance. 
$A_{Z}$ (including $A_1$) affect the model X-ray spectrum\footnote{In the current version (12.13.0c and earlier versions) of XSPEC, for the vvapec model $A_1=1$ independent of the user input value. This is an error which is currently being fixed and will not be in future versions (private communication, K. Arnaud).}. 
Thus in XSPEC, $n_{H,0}$ is the parameter that sets the absolute scale of the number density of atom.

\subsubsection{Electron density, solar abundance}\label{sec:nesol}

The emissivities of different elements which are used in XSPEC are calculated in 
AtomDB\footnote{The AtomDB website  is http://www.atomdb.org/.}  \citep{2012ApJ...756..128F} 
with the XSPEC value of electron density $n_{e,X}=1.2~n_{H,X}$.
 The historical reason for this choice is that the electron density for a fully-ionized plasma with solar abundances is approximately given by this value of $n_e$, and most X-ray emitting plasmas are (approximately) fully-ionized. 

In reality, the electron to hydrogen density $n_{e}/n_{H}$ generally deviates somewhat from 1.2 for a solar abundance plasma. 
Electron densities vary with temperature because the ionization states depend on temperature. 
For the nine choices of solar abundances in XSPEC, and for eight temperatures between 0.1 and 40 keV\footnote{The minimum temperature for the vnei emission models in XSPEC is 0.0808 keV.}, Table~\ref{tab:tab1} gives the ratio $n_{e}/n_H$. 
For solar abundance, i.e. $A_Z=1$ including for hydrogen, thus $n_H=n_{H,0}$; for non-solar abundance XSPEC uses the $n_{H,0}$ to obtain electron density $n_{e,X}$. 
For Table~\ref{tab:tab1} the $vvnei$ model was used with ionization parameter set to $n_e~t=5\times10^{13}$ cm$^{-3}$ s \footnote{Because of the large $n_e~t$ value, the ionization was near-equilibrium.} and we used XSPEC to calculate the ionization states of the elements Z=1 to 30.

The electron density ratios vary with temperature between 0.10 keV and 40 keV by only $\sim$0.2\%. 
The variation of electron to H density between different solar abundance sets is still small, although significantly larger at $\sim$3.5\%.
Thus $n_e=1.2~ n_H$ is an approximation that is accurate to within ~3.5\% for solar abundance plasmas.

\subsubsection{Electron density, non-solar abundance}\label{sec:nenonsol}

 For non-solar abundance, XSPEC assumes $n_{e,X}$ given by Equation~\ref{eqnne}, 
which can be significantly different from the physical value of $n_{e}$. 
To illustrate this, we calculate  $n_{e}/n_{H,0}$ 
for some example sets of abundance factors, $A_Z$. 
The cases are: i) zero H ($A_1$=0); ii) 10\% H and He ($A_1$=$A_2$=0.1);
iii) zero H and He ($A_1$=$A_2$=0); iv) zero H through N ($A_Z$=0 for $Z$=0 to 7);
v) zero H through Al ($A_Z$=0 for $Z$=0 to 13).
For case i) to v), we take elements without a specified abundance factor to have $A_Z$=1. 

A more realistic example uses abundances for the ejecta from a Type Ia supernova explosion. 
The chosen example is one of a set of 3-dimensional hydrodynamic nucleosynthesis calculations for the delayed detonation explosion of a white dwarf carried out by \cite{2013MNRAS.429.1156S}: the N100 model. 
The N100 model has the deflagration ignited in a configuration that results in 0.6 $M_{\odot}$ of Ni56 produced in a 1.4 $M_{\odot}$ CO WD (central density of $2.9\times10^9$ g cm$^{-3}$).
The N100 model has been shown to qualitatively reproduce some key features of spectra and light curves of “normal” Type Ia supernovae reasonably well, 
\citep{2012MNRAS.420.3003S,2012ApJ...750L..19R}, which makes it a good choice as a representative for the class of delayed-detonation explosion models. 
The element masses are specified by mass of each element (Z=1 to 30), with total ejecta mass of 1.4 $M_{\odot}$.
Table~\ref{tab:n100} illustrates the calculation of the abundance factors $A_Z$, with mass of each element given in column 3. 
The element masses are converted to nuclei number using the mass of each nucleus, then normalized so that the number is given relative to number of iron nuclei (Fe is assigned to be 1) in column 4.
The nuclei numbers are converted from relative to iron to relative to solar abundances by dividing by the values of $A_{sol}$, in this case for `angr' abundances, with result given in column 5.
 
To get the $A_Z$ factors typical for a spectral fit, we give two cases.
Column 6 gives $A_Z$ for  $A_{Fe}=1$, and Column 8 gives $A_Z$ for an iron-rich fit with $A_{Fe}$=10. 
Thus $A_Z$ in column 6 or 8 is obtained by dividing the values in column 5 by the factor get the desired $A_{Fe}$ value used to fit the spectrum. 
The chosen value of  $A_{Fe}$ is somewhat arbitrary because all $A_Z$ values can be scaled by an overall scale factor.
This scale factor determines the values of $n_{H,X}$ and affects $n_e/n_{H,0}$. 
As comparison models, we include H and He at solar abundance ($A_1=A_2=1$) with the other abundance factors $A_Z$ in columns 6 and 8 to obtain columns 7 and 9.
These models are added  because models from columns 6 and 7 (or 8 and 9) will produce similar spectra, but have very different electron densities. 
 
For the above examples of abundance factors, and for eight temperatures between 0.1 and 40 keV, the lower half of Table~\ref{tab:tab1} gives $n_e/n_{H,0}$. 
Here we assumed near-equilibrium ionization\footnote{The $vvnei$ model was used with $\tau=5~10^{13}$ cm$^{-3}$s$^{-1}$.} for each temperature to calculate the ionization states of the elements.
For each row (fixed abundance), the electron densities vary with temperature between 0.10 keV and 40 keV by an amount depending on the abundance set. 
Electron density varies very little with temperature for solar abundance (because hydrogen is essentially fully ionized), but for the more metal rich abundance sets it varies signficantly. 
E.g. for zero H and He it varies by 35\%; for zero H to N it varies by 41\%; and for zero H to Al it varies by by 110\%.
This is caused by the heavier elements having more change in ionization states with temperature than lighter elements which are nearly fully ionized even at the lower temperatures.

The variation of the electron density between the different abundance sets is large. 
E.g. consider a temperature of 1.0 keV. 
The electron density for zero H is $\sim$1/6 as much as for solar abundances.
For H and He with $A_i$=0.1, electron density is $\sim$1/10 as much. 
The electron density for zero H and He is $\sim$1/100 as much; for zero H through N it is just slightly below  $\sim$1/100 as much; and  for zero H through Al it is $\sim$1/1000 as much.
Thus the variation in electron density can be large. 

To simplify notation, hereafter we define the electron density ratio as $r_e$, given by:
\begin{equation}
 r_e= \frac{n_e}{1.2~ n_{H,0}} =~\rm{ratio~ of ~electron~ density~ to~ XSPEC~ value} 
 \label{eqnre}
 \end{equation}

\subsection{X-ray spectrum calculation in XSPEC}\label{sec:nion}

The number of photons emitted in a given line $k$ per unit volume per unit time (cm$^{-3}$~s$^{-1}$) from a given ion of charge state q of element, Z, is given by: 
\begin{equation}
 n_{\gamma,k,Z}= n_e ~ n_Z ~ \epsilon_{k,Z} \label{eqnemiss}
 \end{equation}
 with $\epsilon_{k,Z}$ the line photon-emissivity coefficient (cm$^{3}$~s$^{-1}$) multiplied by the ion fraction of that ion, $f(q,Z)=n_q/n_Z$.
To obtain the emission line luminosity (photons s$^{-1}$), one integrates the sum over lines of $n_{\gamma,k,Z}$ over the emission volume, assumed optically thin.
For the case of uniform densities this is given by:
\begin{equation}
\int dV \sum_Z \sum_k n_{\gamma,k,Z} = \int dV \sum_Z \sum_k  n_e ~ n_Z ~ \epsilon_{k,Z}=EM_H~\sum_Z \sum_k~\frac{n_{Z}}{n_{H}} ~ \epsilon_{k,Z}
\label{eqnemis2}
\end{equation}
 where Equation~\ref{eqnEM} has been used.
To obtain the emission line spectrum (photons cm$^{-2}$~s$^{-1}$) at the observer\footnote{Because the energies of the emission lines are known one can bin the spectrum at any step to obtain the spectrum in units of photons cm$^{-2}$~s$^{-1}$~keV$^{-1}$.}, the luminosity is divided by $4\pi D_A^2(1+z)^2$: 
\begin{equation}
\frac{EM_H}{4\pi D_A^2(1+z)^2}~\sum_Z \sum_k~\frac{n_{Z}}{n_{H}} ~ \epsilon_{k,Z}
\label{eqnemis3}
\end{equation}
This yields the model emission line spectrum. 
 
The X-ray spectrum consists of the sum of the emission line photons from all elements which have emission lines in the observed energy range, typically a few tenths of a keV to several keV, plus the continuum photons from electrons and ions in the plasma.
The intensity of continuum photons is proportional to the product of electron and ion densities, so the continuum part of the spectrum scales with electron density and ion density in the same way as the line contribution to the spectrum given by Equation~\ref{eqnemiss}.
Thus the continuum emission can be added as an extra term to the emissivity 
(in units cm$^{-3}$~s$^{-1}$~keV$^{-1}$) in Equation~\ref{eqnemis3}.

XSPEC reads the emissivities per unit electron density per unit hydrogen density precomputed for solar abundances of each element, and summed over all emission lines and all elements  (with an extra factor of $10^{14}$ built in):
 \begin{equation}  
\sum_k ~\frac{n_{Z,sol}}{n_{H,0}} \sum_k ~\epsilon_{k,Z}=\sum_k ~A_{Z,sol} ~ \sum_k ~\epsilon_{k,Z} \label{eqnemis4}
 \end{equation}
XSPEC then adds the continuum emissivity (per unit electron density per unit hydrogen density), and multiplies this by $norm_X$ to obtain the model emission line plus continuum spectrum. 
Thus for solar abundances, the left hand side of Equation~\ref{eqnemis4} times $10^{14}$ times $norm_X$ 
should be the same as Equation~\ref{eqnemis3}.
A direct comparison yields $norm_X$ in terms of $EM_{H,X}$, given by Equation~\ref{eqnXnorm}.

The above is all correct if the abundances are solar. 
When abundances are non-solar, the emissivities are calculated including the abundance factors $A_Z$:
\begin{equation}  
\frac{n_{Z}}{n_{H,0}} \sum_k ~\epsilon_{k,Z}=A_{Z,sol}~ A_Z~ \sum_k ~ \epsilon_{k,Z}  \label{eqnemiss3}
 \end{equation}
 Because the emissivities are multiplied by abundance factors relative to $n_{H,0}$, $norm_X$ obtained from the observed spectrum yields the usual hydrogen and element $Z$ norms from: 
 \begin{equation}  
norm=A_1~norm_X \label{eqnnormC}
  \end{equation}
and
\begin{equation}  
norm_Z=A_Z~A_{sol,Z}~norm_X \label{eqnnormCZ}
  \end{equation}
where we have used Equations~\ref{eqnnH} and \ref{eqnnZ}.
Note that  $norm_X$ is non-zero even if there is no hydrogen ($A_1=0$).
  
$norm_X$ is defined in terms of  $n_{e,X}~n_{H,X}~V/D_A^2$, 
thus Equation~\ref{eqnnormC} yields $n_e~n_H=A_1~n_{e,X}~n_{H,X}=A_1~1.2~n_{H,X}^2$ 
or $n_e~n_{H,0}=n_{e,X}~n_{H,X}=1.2~n_{H,X}^2$ for given $V$ and $D_A$. 
The ionization state and composition gives the ratio $n_e/n_{H,0}$ as discussed above, with examples in Table~\ref{tab:tab1}.
Thus one finds for $n_H$ and other element densities: 
\begin{eqnarray}  
n_{H,0}^2~\frac{n_e}{n_{H,0}} &=& ~ 1.2~n_{H,X}^2~~\rm{or}~~ n_{H,0}=\sqrt{\frac{1.2~n_{H,0}} {n_e}}~n_{H,X}=r_e^{-1/2}~n_{H,X} \label{eqnnH0cor} \\
n_H&=&A_1~r_e^{-1/2}~n_{H,X} \label{eqnnHcor} \\
n_Z&=&A_Z~A_{sol,Z}~r_e^{-1/2}~n_{H,X} \label{eqnnZcor}
\end{eqnarray}
with $n_{H,X}$ found from  the definition of $norm_X$ (Equation~\ref{eqnXnorm}) given $V$ and $D_A$\footnote{Here we take redshift $z=0$, noting that for non-zero redshift, one replaces $D_A$ by $D_A~(1+z)$.}:
 \begin{equation}  
1.2~n_{H,X}^2=10^{14}~norm_X~\frac{4~\pi~D_A^2}{V} \label{eqnnHX}
\end{equation}
Thus the fiducial density is given in terms of $norm_X$ by:
 \begin{eqnarray}  
  n_{H,0}&=&r_e^{-1/2}~\sqrt{10^{14}~norm_X~\frac{4~\pi~D_A^2}{1.2~V}}
 \label{eqnnH0f}
\end{eqnarray}

\subsubsection{Constrained and unconstrained elements}\label{sec:spec}   

Here we make a distinction between the elements with abundances constrained by a spectral fit to the observed spectrum and those elements  
which do not contribute to the observed spectrum\footnote{I.e., they do not have significant emission line emission or continuum emission in the observed energy range.}. 
Normally their abundance factors are only constrained to be below a certain upper limit, which can be large.
The upper limit can be determined by fitting an observed X-ray spectrum with increased abundance factor of that element until the spectral fit becomes statistically poor.  

\begin{figure*}[ht!]
\plotone{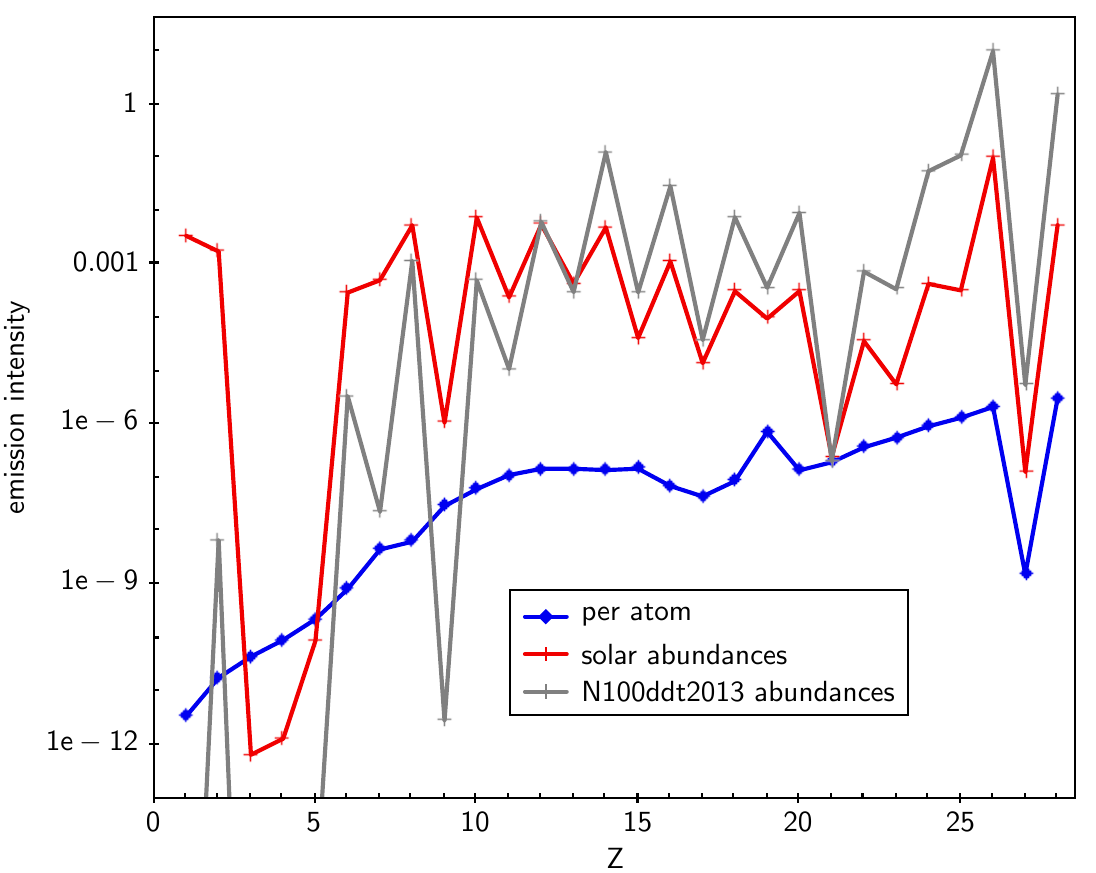}
\caption{Illustration of the relative contributions of different elements to the 0.3 to 8 keV energy range for a spectrum, using the response matrices for the AstroSat SXT instrument \citep{2017JApA...38...29S}. The maximum intensity in photons s$^{-1}$ keV$^{-1}$ is shown for elements $Z=1$ to 28 for three cases: per atom, for solar abundances and for N100ddt2013 abundances. For each case the vertical scale is arbitrary. The XSPEC vvapec model was used with temperature of 1 keV. For N100ddt2013, H, Li, Be and B not shown because they are so small: they at $2.9\times10^{-22}$,  $1.9\times10^{-24}$, $4.1\times10^{-33}$,  and $1.2\times10^{-16}$ respectively times the contribution of Fe.}
\label{fig:intens}
\end{figure*}

For illustration, we used XSPEC\footnote{For this calculation the spectral response of the AstroSat SXT instrument \citep{2017JApA...38...29S} was used, over the energy range of 0.3 to 8 keV and the vvapec model was used with a temperature of 1 keV.}  
to calculate the peak intensity from a plasma consisting of a single element. 
This was done for the elements Z=1 to 28 and the results are shown in Figure~\ref{fig:intens}.
Three different curves are shown, where each curve is given its own normalization, for comparison of the contributions of different elements to the X-ray spectrum.
The blue curve is the intensity normalized per atom, and ranges from a minimum for H to a maximum for Ni. 
Thus one atom of Ni contributes as much 0.3-8 keV X-ray emission as $8.9~10^5$ atoms of H. 

The shape of the spectrum for an astrophysical plasma is determined by a combination of abundances and the intensity of the emitted spectrum per atom. 
Solar abundances are dominated by H and He and deficient in Li, Be and B. 
The resulting X-ray emission from the different elements for solar abundance is shown by the red curve in Figure~\ref{fig:intens}, with the very small contribution from Li, Be, B. 
The dominant contribution is from Fe, with O, Ne Mg, Si and Ni each at 0.05-0.08 the contribution from Fe. 
H and He are next with contributions .034 and 0.017 times that of Fe. 
The grey curve is for the N100 model abundances, which are deficient in H through B as well as N and F. 
In this case Fe and Ni (at 0.15 times that of Fe) are the strongest contributors to the spectrum. 
Smaller contributions come from the elements Si and Mn with 0.01 times the contribution of Fe.
For the N100 model, the elements  $Z$=1 to 12 contribute little to the X-ray spectrum (by factors relative to Fe of $\sim10^{-3}$ for O to $\sim10^{-32}$ for Be). 
in particular H and He contribute  $2.9~10^{-22}$ and  $6.8~10^{-10}$ as much as a Fe.
 
\begin{figure*}[ht!]
\plotone{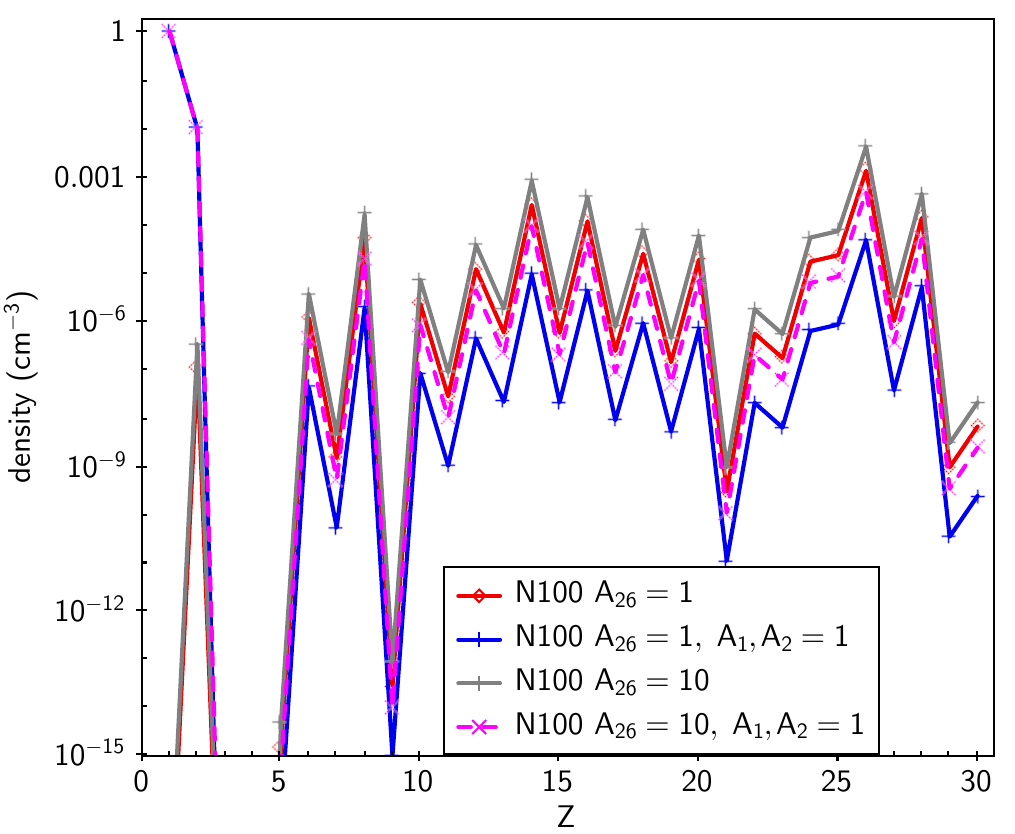}
\caption{Element densities for N100 model \citep{2013MNRAS.429.1156S}, with and without added H and He. 
For calculation, the distance is 1 kpc, the emitting volume is a sphere of radius 2 pc, and the XSPEC norm is 0.1,
which yields XSPEC $EM_X=1.2\times10^{57}$ cm$^{-3}$.
The four cases shown are N100 abundances normalized with $A_{Fe}$=1, N100 with  $A_{Fe}$=1 and H and He added at solar abundances,   N100 abundances normalized with $A_{Fe}$=10, and N100 with  $A_{Fe}$=10 and H and He added at solar abundances.
}
\label{fig:abund2}
\end{figure*}

For illustration we estimate an element abundance to be unconstrained if its maximum contribution is less than 1\%\footnote{The ability to distinguish the contribution of a given element can vary greatly. It is easier to distinguish line emission if the strongest lines from an element do not overlap those from other more strongly emitting elements, and it can be difficult to distinguish smooth continuum such elements such as from H and He which have no lines in the X-ray range.} as that from Fe, 
with $A_{26}$ set to 1\footnote{If $A_{26}$ is set to 10, the quoted values below for a 1\% contribution relative to Fe are all increased by a factor 10.}.
Figure~\ref{fig:abund2} shows the $A_Z$ values (blue line) giving this 1\% of Fe. 
The maximum abundance relative to solar can be quite large for the rare elements Li, Be and B:
 they are unconstrained if the abundance factors $A_Z$ are less than   $1.5~10^{9}$,  $7.8~10^{8}$,  and $1.1~10^{7}$, resp.
F, P, Cl, K, Sc, Ti, V and Co are unconstrained if their $A_Z$ are less than  $930$,  $24$,  $70$,  $10$, $4000$,  $27$, $180$ and $7800$, resp.
The remainder of elements are unconstrained if their $A_Z$ are less than 0.1 to 3, in particular for  H if $A_1<0.3$ and for He if $A_2<0.6$.
The spectra and their intensities depend on plasma temperature, so there will be some variation in these upper limits with temperature.

The issue of which abundances can be modified (or have upper limits) while retaining a satisfactory fit to the X-ray spectrum should be determined using XSPEC. 
The purpose of the above discussion was to provide guidelines to determine which elements are unconstrained elements.
For elements with upper limits, one can choose any value for their abundances below their upper limits. 
Typically, the abundances of unconstrained elements are based on theoretical considerations of the composition of the shock heated plasma.
E.g. typical interstellar medium in the disk of the Galaxy is expected to be of near solar abundance, or the composition of the ejecta of a supernova remnant can be estimated from nucleosynthesis calculations for a stellar explosion consistent with that remnant.

In summary, one has two sets of elements:  constrained and unconstrained.
The constrained elements have their $A_Z$ values determined by an XSPEC spectral fit and the unconstrained elements have their $A_Z$ values chosen using other considerations, with upper limits consistent with the XSPEC spectral fit. 
Here we label the unconstrained element abundance factors $A_{Z,u}$ and constrained element abundance factors $A_{Z,c}$.

\subsubsection{H and He abundances from simulated X-ray spectra}\label{sec:spec} 

As illustrated in Figure~\ref{fig:intens}, H and He have weak emission per atom compared to heavier elements.
However given that they dominate solar abundances, they are important for hot plasma spectra of solar abundance.
For heavy element rich plasmas, the emission from H and He is weak (e.g. for N100 abundances shown in Figure~\ref{fig:intens} by the grey line).

To determine the extent to which H and He are unconstrained elements, we created simulated 0.5 to 7 keV X-ray spectra with different signal-to-noise which are typical of supernova remnants in the Galaxy.
In particular we consider the reverse-shock emission from Type Ia supernova remnants (SNRs), which should represent most closely the element abundances of the N100 model.  
A list of Galactic SNRs with reverse shock emission detected is given in Table 1 of \citet{2020ApJS..248...16L},
with model ages and interstellar medium densities given in Table 2 of that paper.
The data extracted from  \citet{2020ApJS..248...16L} for the Type Ia SNRs is given in Table~\ref{tab:snrem} below,
as well as the mean parameters needed for a spectral fit in XSPEC (norm, kT and $n_e t$).
The average Type Ia in the Galaxy has reverse shock spectrum with norm\footnote{$norm_X$ has units of $10^{-14}$ cm$^{-5}$, which are normally omitted.} of $2.32\times 10^{-3}$, kT=2.60 keV, and 
$n_e t=1.31 \times10^{12}$ cm$^{-3}$ s. 

We give results here for three simulated X-ray spectra. 
These use the standard rmf and arf files for the SXT instrument on AstroSAT and the standard SXT background spectrum (see the AstroSAT Science Support Cell at http://astrosat-ssc.iucaa.in/ for details). 
The first simulated spectrum uses an exposure time of 10 ks and the average Galactic Type Ia reverse shock parameters listed in Table~\ref{tab:snrem}. 
We call this the low signal-to-noise spectrum (LSN).
The second spectrum uses an exposure time of 20 ks and a norm 3 times higher (same  kT and $n_e t$ as LSN)
to obtain a medium signal-to-noise spectrum (MSN).
The third spectrum uses a long exposure time of 40 ks and a norm 10 times higher (same  kT and $n_e t$ as LSN) to obtain a high signal-to-noise spectrum (HSN).
The input element abundances for LSN, MSN and HSN cases were the same: the N100 values from column 8 Table~\ref{tab:n100} (the $A_{Z,Fe10}$ values) and input column density $N_H$ was $10^{21}$ cm$^{-2}$. 
The three spectra were rebinned to obtain a minimum 10 counts per bin.

The LSN is typical for an observation with AstroSAT SXT of reverse shock emission from Galactic Type Ia SNR. MSN is twice as bright (twice the norm) as the brightest observed SNR; and HSN is five times brighter than the brightest observed SNR.
However AstroSAT SXT \citep{2017JApA...38...29S} has an effective area of $\sim$100 cm${^2}$ over 0.5 to 7 keV, compared to values of $\sim$500 cm${^2}$, $\sim$300 cm${^2}$, $\sim$1000 cm${^2}$ and $\sim$1000 cm${^2}$ for Chandra ACIS (https://cxc.harvard.edu/), XMM Newton MOS and pn (https://www.cosmos.esa.int/web/xmm-newton/technical-details-epic), and Suzaku XIS(https://space.mit.edu/XIS), respectively.
Thus observations of Galactic Type Ia SNR with Chandra, XMM Newton and Suzaku would have signal to noise for reverse shock emission similar to the MSN or HSN cases. 
The spectral resolution of SXT is 80 to 150 eV over the 0.3 to 8 keV band \citep{2021JApA...42...17B}.
This is comparable to the instruments above\footnote{The spectral resolutions, quoted at 2 keV from the instrument websites listed above, are 70 eV (ACIS FI), 130 eV (ACIS BI), 80 eV (MOS), 120 eV (pn), and 75 eV (XIS).}. 

\begin{figure*}[ht!]
\plottwo{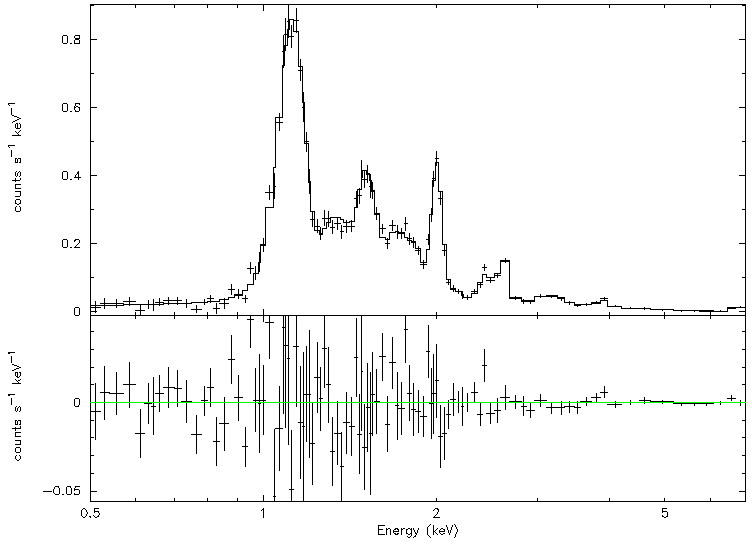}{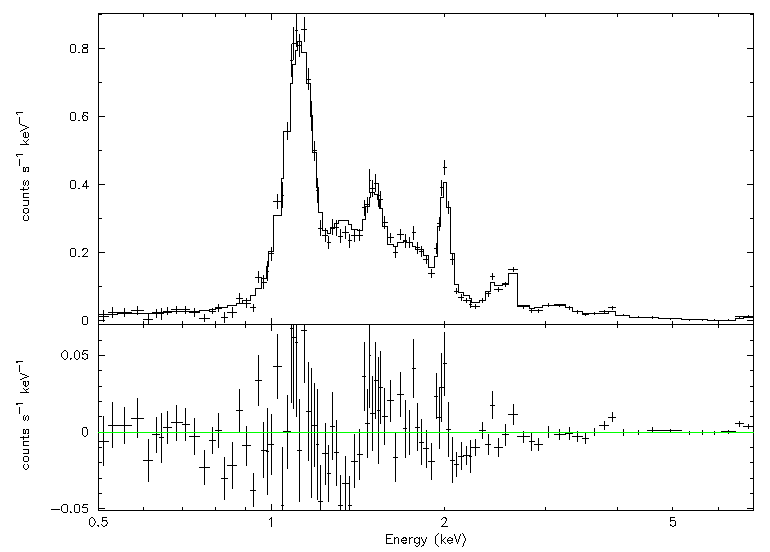}
\caption{The HSN (high signal-to-noise, see text) simulated X-ray spectrum, which uses rmf, arf and background files for AstroSAT SXT, and abundances given by column 8 in Table~\ref{tab:n100} and model fits. The other spectrum parameters are $N_H=10^{21}$ cm$^{-2}$, kT=2.6 keV, $n_e~t=1.31\times10^{12}$ cm$^{-3}$ s and norm=$2.32\times10^{-2}$. The fit model on the left has original N100 abundances, whereas the fit model on the right has added H and He with $A_{H}=A_{He}=1$. Although the fit with solar H and He is statistically poor (probability=1.6E-06, Table~\ref{tab:sim}), it is not clearly different except for larger residuals by a factor of $\sim$1.2.
}
\label{fig:simspec}
\end{figure*}

The three spectra (LSN, MSN, HSN) were fit using XSPEC with all abundances except H and He frozen to the
values from Table~\ref{tab:n100} and fit with free parameters $N_H$, kT, $n_e t$ and norm.
Then we tested how much H and He can be added until the fits become statistically unacceptable.
Figure~\ref{fig:simspec} (left) shows the HSN spectrum and model fit with original small amount of H and He in the N100 abundances (column 8 in Table~\ref{tab:n100}). 
The fit with add H and He at solar abundance is shown on the right. 
The fit on the right is statistically rejected, with probability $1.6\times10^{-6}$, but is not visibly different except that it has higher residuals.
We tested freeing the abundances of various elements and found that freeing several elements improved the fit significantly but not enough to be statistically acceptable. 
E.g. making Fe and Ni abundances free increased probability to $1.9\times10^{-2}$.
To obtain a statistically acceptable fit it was necessary to make H or He free parameters.
In the case with original N100 abundances except for H and He, with the fit with H=1 and free He gave He abundance $0\pm0.8$ and probability 0.068, 
and the fit with He=1 and free H gave H abundance $0\pm0.26$ and probability 0.23.
Thus one could conclude that the model with H=1 and He=1 is statistically ruled out, but some significant H and He are allowed in the spectral fit (up to $\sim0.5$ times solar).

\section{Results and discussion}\label{sec:results}  \label{sec:results} 
 
\subsection{Emission measure}\label{sec:EM}

XSPEC yields $norm_{X}$ which can be converted to $EM_{H,X}$ using $V$ and $D_L$ 
from the spectral fit to the observed spectrum\footnote{Here we assume the user has entered the desired $A_{Z,u}$ for unconstrained elements and has used the spectral fit to obtain $A_{Z,c}$ for the constrained elements.}.
$norm_{X}$ is always non-zero because it is based on a fiducial value of $n_{H,0}$ with the actual value given by $n_{H}=A_1~n_{H,0}$. 
$norm_{X}$ can be converted to $norm$ (for H) or $norm_{Z }$ by multiplying by the ratio of element density to the fiducial hydrogen density, which is $A_1$ for hydrogen and $A_Z~A_{sol,Z}$ for other elements (Equations~\ref{eqnnormC} and \ref{eqnnormCZ}). 

\subsection{Element densities corrected for electron density}\label{sec:nZ}  

The electron density is computed from the element abundance factors $A_{Z,c}$ and $A_{Z,u}$, $Z=1$ to 30, and the ion fractions for each element, which can be computed using XSPEC\footnote{If one sets the verbose level in XSPEC to 25, then XSPEC outputs the ion fractions for each element in the log file.}.
From the ionization states and element abundance factors for elements $Z=1$ to 30, one calculates the electron density:
\begin{equation}
n_e=n_{H,0}~\sum_{Z=1}^{30}  A_{sol,Z} ~ A_Z \sum_{i=1}^{Z}i~ f_{i,Z}
\end{equation}
with $f_{i,Z}$ the fractional abundance of ion state $i$ for element $Z$ for the given plasma conditions (e.g. temperature and ionization age).
Table~\ref{tab:tab1} gives several examples of calculation of electron densities.
Thus one can obtain the ratio $n_e/n_{H,0}$.

To get physical values of $n_e$ in terms of $norm_X$, one uses $n_{H,0}$ from Equation~\ref{eqnnH0f}:
\begin{eqnarray}
n_{e}&=&\frac{n_e}{1.2~n_{H,0}} ~ 1.2~n_{H,0} =1.2~r_e~n_{H,0}  \nonumber \\ 
            &=&1.2~  r_e^{1/2}~\sqrt{10^{14}~norm_X~\frac{4~\pi~D_A^2}{1.2~V}}
\label{eqnneCXX}
 \end{eqnarray}
 Similary $n_H$ and $n_Z$ in terms of $norm_X$ are: 
 \begin{eqnarray}
 n_{H}&=&A_1~r_e^{-1/2}~\sqrt{10^{14}~norm_X~\frac{4~\pi~D_A^2}{1.2~V}} \nonumber \\
 n_{Z}&=&A_Z~A_{sol,Z}~r_e^{-1/2}~\sqrt{10^{14}~norm_X~\frac{4~\pi~D_A^2}{1.2~V}}
\label{eqnnHZf}
 \end{eqnarray}
 
 One can see from Equation~\ref{eqnnHZf} that the correct value of $n_H$ is
  larger by the factor $A_1~r_e^{-1/2}$ than $n_{H,X}$.
Because of the $A_1$ factor, this factor is usually less than 1 for H-poor plasmas. 
The correct values of $n_Z$ are larger for elements enriched in the plasma by the factor $A_Z~r_e^{-1/2}$ than $A_{sol,Z}~n_{H,X}$. 
For depleted elements ($A_Z<1$) the density can be larger or smaller than the XSPEC value depending on whether $A_Z$ is larger or smaller than $r_e^{1/2}$.
The electron density is smaller than the XSPEC value $1.2~n_{H,0}$ by factor $r_e^{1/2}$.
 
For unconstrained elements, the corrected densities are given by the user specified abundances $A_{Z,u}$.
Normally these are anchored to the constrained elements, with one constrained element $Z0$ chosen as the anchor (e.g. iron):
\begin{equation}
n_{Z,u}= \frac{n_{Z,u}}{n_{Z0,c}}  ~n_{Z0,c} \label{eqnCoru}
 \end{equation}

\begin{figure*}[ht!]
\plotone{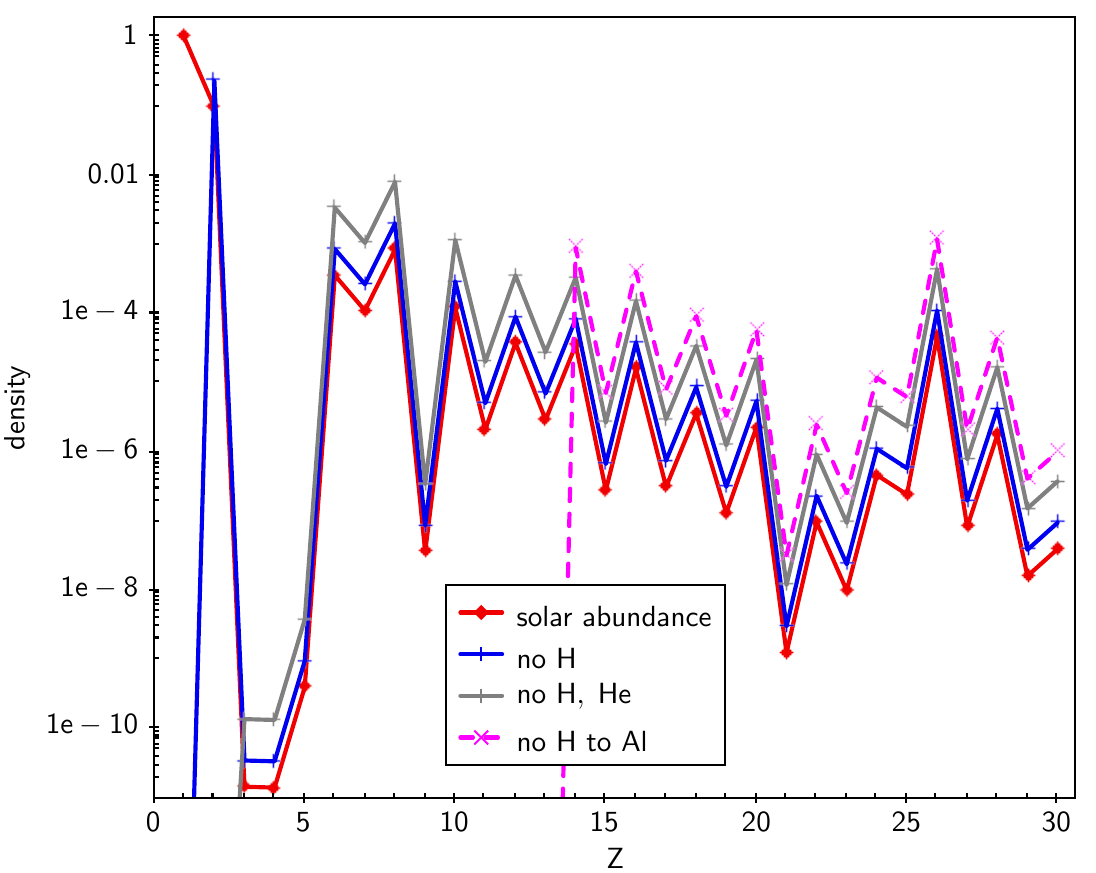}
\caption{Element densities for different cases. 
The four cases shown are solar abundance, no H, no H and He, and no H through Al.
}
\label{fig:abund}
\end{figure*}

As an example calculation of element densities, we use the four cases from line 1 (solar cases) and lines 1, 3 and 5 (non-solar cases) of Table~\ref{tab:tab1} with plasma temperature of 2 keV. 
We choose $D_A$ and $V$ in the emission measure to give $n_H$=1 cm$^{-3}$ for the solar abundance case.
 The calculated element densities, $n_{Z}$,  are plotted in Figure~\ref{fig:abund}.
 The `no H' case has $\simeq$2.4 times the densities of elements $Z=2$ to 30 as the solar abundance case (shown in black). 
The `no H, He' case has $\simeq$10 times the densities of elements $Z=3$ to 30 as the solar abundance case, and the
 `no H to Al' case has $\simeq$26 times the densites of elements $Z=14$ to 30 as the solar abundance case.
  One can see from Figure~\ref{fig:abund}, that the densities of all elements are sensitive to the composition assumed.
  
As a more realistic case of a light element deficient plasma we use the N100 abundances for SN Type Ia ejecta, as described above. 
For N100 abundances, we consider 4 cases.
The calculated abundances for all $A_Z$ with no adjustment to the unconstrained element abundances which yield:
case i with $A_{26}=1$; case ii the same as case i but with added H and He with $A_1=A_2=1$;
case iii with $A_{26}=10$ (i.e. all $A_Z$ a factor of 10 higher than case i);  case iv the same as case i but with added H and He with $A_1=A_2=1$.
For case ii the added H and He are above the 1\% contribution relative to Fe (at 3.3\% and 1.6\% respectively, but for case iv the added H and He are below the 1\% contribution relative to Fe (at 0.33\% and 0.16\% respectively). 
The model spectra for all for cases are similar and representative of an observed N100 spectrum\footnote{This depends on the detector sensitivity and resolution and the signal to noise of the observed spectrum.}.

Figure~\ref{fig:abund2} shows the calculated element densities for the 4 cases for a plasma with temperature of 2 keV. 
The heavy element densities for the models without added H and He are higher by a factors of $\simeq$28 (case i higher than case ii) and $\simeq$8.8 (case iii higher than case iv). 
This is a direct result of the difference in electron density between cases i and ii and between cases iii and iv, given in Table~\ref{tab:tab1}.

\subsection{Element masses}\label{sec:Zmass}

Element masses depend on the element densities and emitting volume,
\begin{eqnarray}
M_Z&=&m_Z \int n_Z(\vec{r}) dV=m_Z~n_Z ~V~\rm{(if~uniform)} \nonumber \\
&=&m_Z~EM_Z /n_e =m_Z~A_{sol,Z}~A_Z~EM_{H,X}/n_{e}  \nonumber \\
&=&m_Z~A_{sol,Z}~A_Z~\sqrt{\frac{10^{14}~norm_X~4~\pi~D_A^2~V}{1.2~ r_e}}
\label{Mz}
 \end{eqnarray}
 with $m_Z$ the mass of one atom\footnote{The mass of element $Z$, $m_Z$, depend on the relative abundances of the different isotopes.} of element $Z$, $EM_Z$ given by Equation~\ref{eqnnormCZ} and $n_e$ by Equation~\ref{eqnneCXX}. 
 Element masses are larger than what is inferred from the XSPEC norm by the factor $r_e^{-1/2}$, or $A_Z ~ r_e^{-1/2}$ larger than for the XSPEC solar abundance case\footnote{As in Section~\ref{sec:EMgen}, for the non-uniform case we can use
  $n_Z(\vec{r}) = g_Z(\vec{r})n_Z$ with normalization of $g_Z(\vec{r})$ given by 
$V=\int g_Z(\vec{r}) dV$.}.

\begin{figure*}[ht!]
\plotone{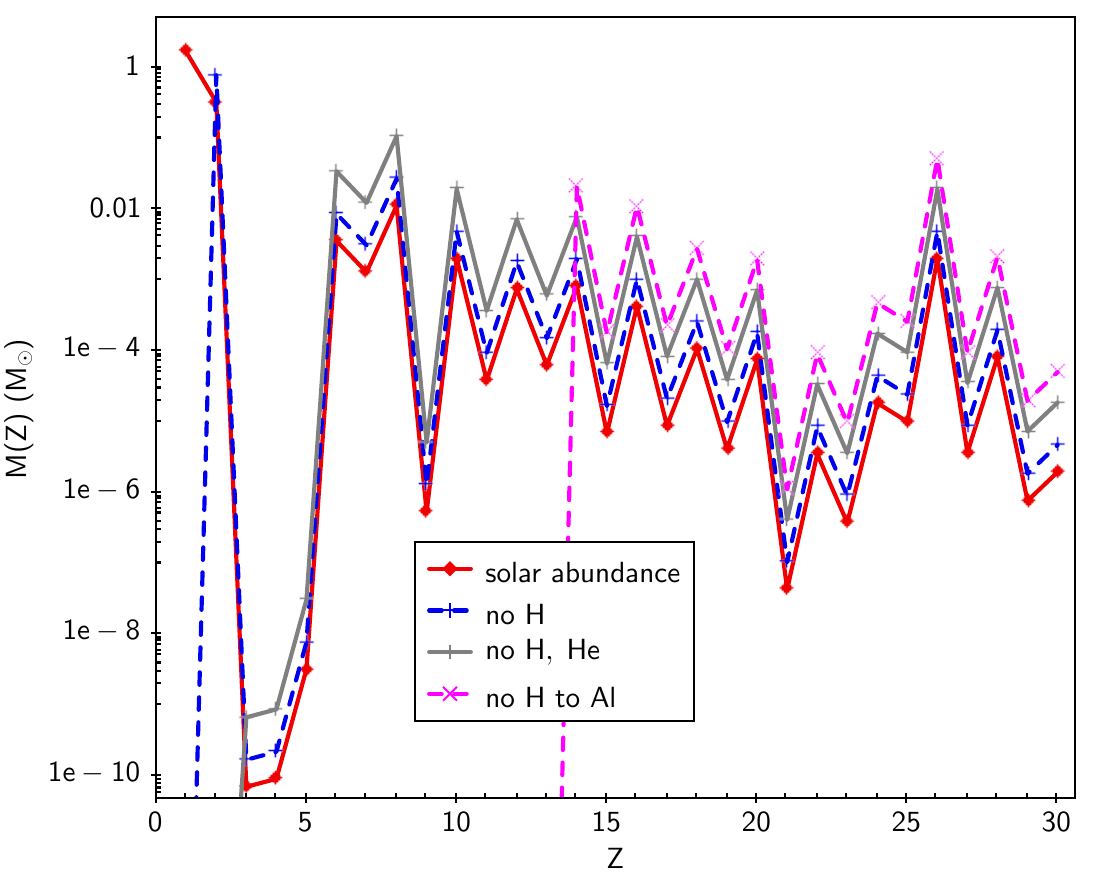}
\caption{  Masses for different cases. 
For calculation, the distance is 1 kpc, the emitting volume is a sphere of radius 2 pc, and the XSPEC norm is 0.1.
The four cases shown are solar abundance, no H, no H and He and no H through Al. 
}
\label{fig:mass}
\end{figure*}

The chosen example has an emitting plasma is a sphere of radius 2 pc with distance 1 kpc and  XSPEC $norm_X=0.1$, which yields XSPEC $EM_X=1.2\times10^{57}$ cm$^{-3}$. 
Figure~\ref{fig:mass} shows masses of elements $Z=1$ to 30 for solar abundances and the no H, no H and He, and no H through Al cases. 
One can see that the masses for the non-solar abundance cases are significantly larger than the solar abundance case, even though the emission measures $EM_Z$ are the same as the XSPEC values.
E.g. for iron, the no H case has 2.4 times larger mass, the no H,He case has 9.5 times larger mass
and the no H to Al case has 26 times larger mass.

\begin{figure*}[ht!]
\plotone{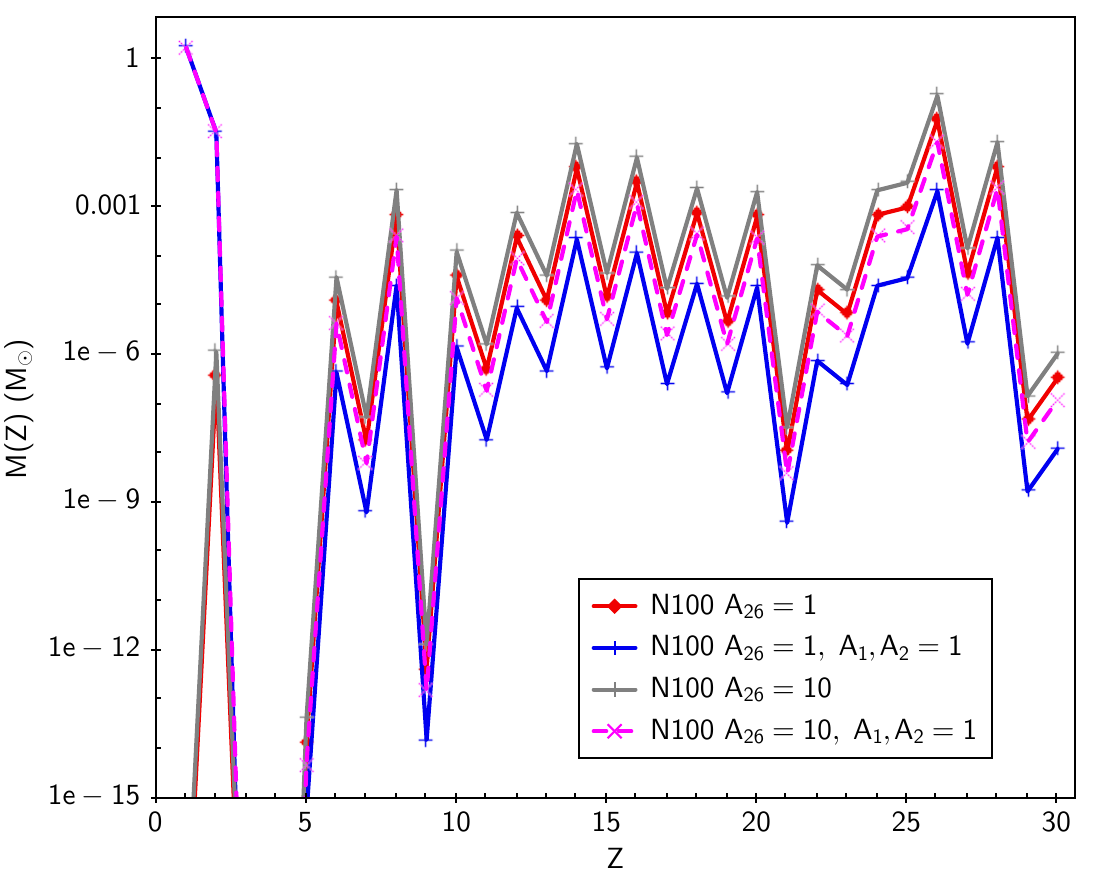}
\caption{Masses for different cases. 
For calculation, the distance is 1 kpc, the emitting volume is a sphere of radius 2 pc, and the XSPEC norm is 0.1.
The four cases shown are N100 abundances normalized with $A_{Fe}$=1, N100 with  $A_{Fe}$=1 and H and He added at solar abundances,   N100 abundances normalized with $A_{Fe}$=10, and N100 with  $A_{Fe}$=10 and H and He added at solar abundances.
}
\label{fig:massn100}
\end{figure*}

For the Type Ia N100 abundances, Figure~\ref{fig:massn100} shows masses of elements $Z=1$ to 30 comparing the four cases of N100 abundances defined above. 
The masses for the standard N100 abundance cases are significantly larger than for the N100 abundances with added H and He. 
For the N100 case with $Z_{26}=1$, the mass is 28 times higher than the case with added H and He, and for N100 with $Z_{26}=10$, the mass is 8.8 times higher than the case with added H and He.
Because $norm_X$ is tied to the fiducial hydrogen density, the N100case with $Z_{26}=10$ has $\sqrt{10}$ times higher mass of Fe than the  N100 case with $Z_{26}=1$.
 
\subsection{H and He as unconstrained elements for heavy element-rich spectra}\label{sec:HHe} 

Tests were carried out for different simulated spectra using N100 abundances and different levels of signal-to-noise (Section~\ref{sec:spec}) to determine how much H and He could be added to the model fits before they disagree with the input simulated spectrum.
For each fit, the parameters $N_H$, kT, $n_e t$ and norm were free and all abundances were frozen, with $A_Z$ ($Z\ne1$ or 2) at the N100 values.

The results for the fits to the three different signal-to-noise spectra (LSN, MSN and HSN) and different amounts of H and He are given in Table~\ref{tab:sim}.
For the LSN case, the added H and He can be up to $\sim$20-30 times solar before the fits become clearly rejected (probability $\sim$10 times lower than for true abundances, with essentially no H and He).
For the MSN case, the added H and He can be up to $\sim$2 times solar before the fits become clearly rejected.
The HSN case is most restrictive on the amount of  added H and He:
it can be up to $\sim$0.5 times solar before the fits become clearly rejected.

This can be interpreted as follows: for low signal-to-noise spectra, H and He are unconstrained up to $\sim$20 times solar, for medium signal-to-noise spectra, H and He are unconstrained up to $\sim$2 times solar,
and for high signal-to-noise spectra, H and He are unconstrained up to $\sim$0.5 times solar.
The details of the upper limits to H and He depend in detail on the observed spectrum exposure time and resolution.
However the tests above are indicative of what level of unconstrained abundance for H and He are for current X-ray astronomy instruments (including Chandra ACIS, XMM MOS and PN, Suzaku XIS and AstroSAT SXT).

The implication is that the variation in electron density ratios ($n_e/n_{H,0}$) given in Table~\ref{tab:tab1} are realistic, and the large corrections for H and He poor spectra, shown in Figures~\ref{fig:abund}, \ref{fig:abund2}, \ref{fig:mass} and \ref{fig:massn100} may be necessary.

\section{Summary and Conclusion}\label{sec:sum}

XSPEC makes specific assumptions to calculate emission measure which are not accurate for non-solar abundances. 
Here we explain the assumptions and generalize to the case of non-solar abundances to obtain exact expressions for emission measures and for element densities and masses derived from emission measure. 
Previous work by \cite{2022Univ....8..274L} developed an approximate correction to emission measures for non-solar abundances.
The results of the current work are recommended for use instead of the earlier approximate results.

Hot plasma spectral model fits (such as $vvnei$) in XSPEC yield the emission measure $EM_{H,X}$ (given by  Equation~\ref{eqnXnorm}) and yield abundance factors $A_Z$.
The main complication with interpretation of results from XSPEC is that the electron density $n_{e,X}$ is given by Equation~\ref{eqnne}. 
The hydrogen density used to fit the X-ray spectrum is $n_H$ which is different than
$n_{H,0}$ if $A_1\ne1$.
The true electron density $n_e$ can be different than $n_{e,X}$ depending on the plasma composition and temperature.
However, $n_e$ can be calculated for any given composition/temperature using XSPEC.

In the case of solar abundances, the electron density can be up to $\sim3.5$\% different than that given by Equation~\ref{eqnne}, as illustrated by Table~\ref{tab:tab1}.
The electron density for solar abundances is dominated by H ($\simeq$83\%) and He ($\simeq$17\%), with all other elements combined contributing $\simeq$1\% to the electron density.
For Equation~\ref{eqnne} to be a good approximation to $n_e$, one must also normalize solar abundances with abundance factors $A_1$=1 and $A_2$=1.

For cases of non-solar abundances, with $A_1$ or $A_2$ different from 1, significant corrections are needed to the element densities to account for the difference between $n_e$ and $n_{e,X}$.
$n_e/n_{e,X}$ can be much less than 1 as illustrated by the examples in Table~\ref{tab:tab1}. 
We define the electron density ratio $r_e$ (Equation~\ref{eqnre}).
The  corrected densities are higher, by the factor $r_e^{-1/2}$ as given by Equation~\ref{eqnnHZf}.
The inferred element masses are corrected by the same factor ($r_e^{-1/2}$) thus can also be much larger, given by Equation~\ref{Mz}.
This implies that masses that may have been derived in the past using $norm$'s from XSPEC spectral fitting on hydrogen poor plasmas, such as expected for Type Ia SNRs, may be significantly underestimated.

\acknowledgments
This work was supported by a grant from the Natural Sciences and Engineering Research Council of Canada.

\clearpage



\begin{figure*}[ht!]
\plotone{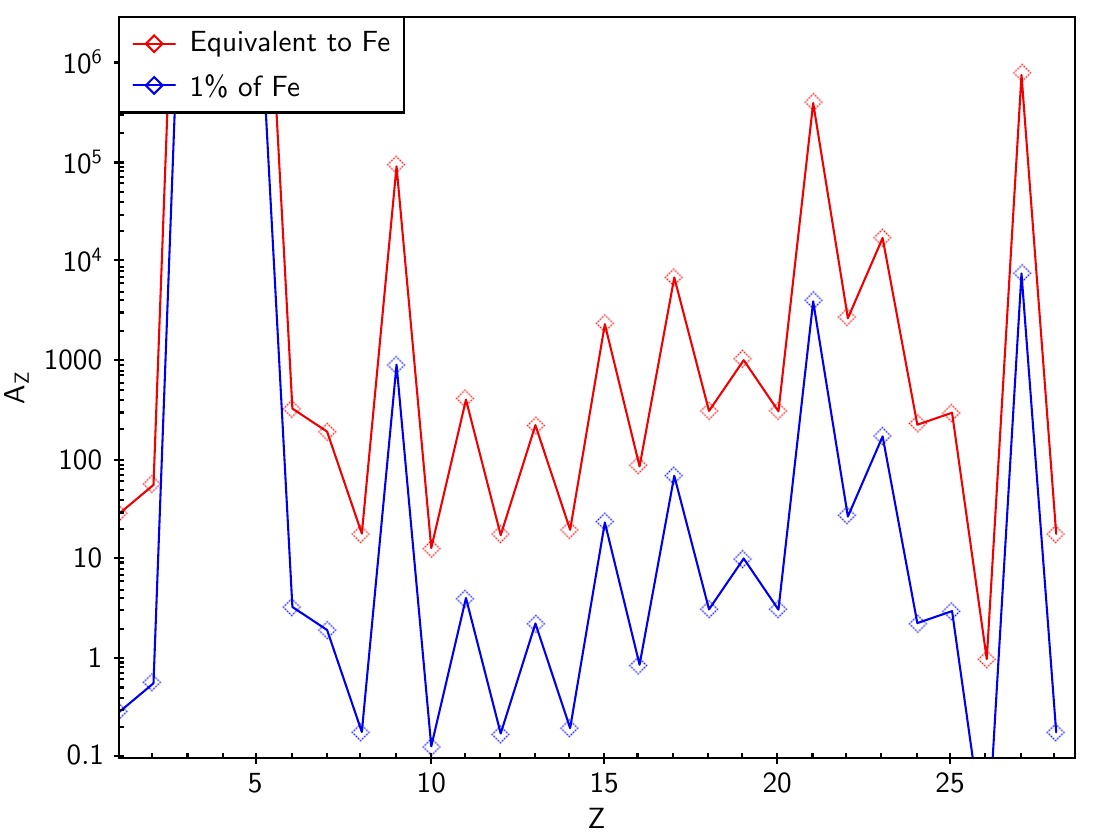}
\caption{The abundance factors $A_Z$ required to yield the same maximum contribution (red line) or to yield a contribution 1\% as large (blue line) as the contribution from Fe  to the 0.3 to 8 keV X-ray spectrum. The response matrices for the AstroSat SXT instrument was used and an equilibrium spectrum with temperature kT=2 keV and the abundance factor of Fe was set to 1 for the same-contribution line (or 0.01 for the 1\% line). 
The abundance factors for Li, Be and B off scale, at values of $1.5\times10^{11}$,  $7.8\times10^{10}$,  and $1.1\times10^{9}$ for equal contribution, with values 0.01 times as large for a 1\% contribution.}
\label{fig:maxAZ}
\end{figure*} 

\clearpage

\clearpage

\begin{deluxetable*}{cccccccccccc}
\tabletypesize{\scriptsize}
\tablecaption{Electron-to-(fiducial)hydrogen density ratio $n_{e}/n_{H,0}$ for different cases of  composition and different temperatures.} 
\tablewidth{0pt}
\tablehead{\\ 
 \colhead{ kT(keV):} & \colhead {0.10} & \colhead {0.25} &  \colhead{0.50}   &   \colhead{1.0}   &   \colhead{2.0}  & \colhead{4.0}   &   \colhead{10.0}  &   \colhead{40.0} \\
 \colhead{COMPOSITION}  
}
\startdata
SOLAR CASES$^{a}$\\
angr 	&1.20532 & 1.20652	&  1.20769 &1.20836& 1.20866 & 1.20880 &1.20885  &1.20886\\
aspl		&1.17672 & 1.17756	&  1.17829 &1.17871& 1.17893 & 1.17903 &1.17907  &1.17908\\
feld	&1.20536 & 1.20647	&  1.20765 &1.20829& 1.20856 & 1.20867 &1.20872  &1.20873\\
aneb &1.16981 &	1.17092	&  1.17198 &1.17258& 1.17285 & 1.17297 &1.17302  &1.17303\\
grsa  &1.17861 & 1.17962	&  1.18058 &1.18113& 1.18138 & 1.18149 &1.18154  &1.18155\\
wilm &1.20152 & 1.20224	&  1.20295 &1.20335& 1.20354 & 1.20362 &1.20365  &1.20366\\
lodd  &1.16470 &	1.16552	& 1.16625 &1.16666& 1.16686& 1.16696&1.16670 &1.16670\\
lpgp  &1.17506 & 1.17589	&  1.17668 &1.17713& 1.17736 & 1.17746 &1.17775 &1.17775\\
lpgs  &1.20159 & 1.20254	&  1.20342 &1.20394& 1.20420 & 1.20431 &1.20436  &1.20437\\
NON-SOLAR CASES\\
 H$~$0$^{b}$	& 0.20532 & 0.20652	&  0.20769 &0.20836& 0.20866 & 0.20880 &0.20885  &0.20886\\
(H,He)$~$0.1$^{b}$	&	0.12946 & 0.13066	&  0.13183 &0.13250& 0.13280& 0.13294 &0.13299  &0.13300\\
 (H,He)$~$0$^{b}$ &0.00992 &	0.01112	&  0.01229 &0.01296&0.01326 &0.01340 &0.01345  &0.01346\\
 (H to N)$~$0$^{b}$  & 	0.00743 &0.00833	&  0.00935 &0.01000&0.01030 & 0.01043 &0.01049 &0.01050\\
 (H to Al)$~$0$^{b}$  &0.00093 & 0.00142	&  0.00155 &0.00168& 0.00183 & 0.00192&0.00195  &0.00196\\
N100Fe1$^{c}$  &0.00108 & 0.00157	&  0.00172 &0.00186& 0.00202 & 0.00211&0.00215  &0.00215\\
N100Fe1HHe$^{d}$  &1.1965 & 1.1970	& 1.1971 &1.1973&1.1974& 1.1975&1.1975  &1.1976\\
N100Fe10$^{e}$  &0.01080 & 0.01574	&  0.01720 &0.01862& 0.02021& 0.02112&0.02150  &0.02152\\
N100Fe10HHe$^{f}$  &1.2062 & 1.2111 & 1.2126	& 1.2140& 1.2156&1.2165&1.2169  &1.2169\\
\enddata
\tablenotetext{a}{Abundance references given in the XSPEC manual- see the abund command.}
\tablenotetext{b}{All other elements taken to have solar abundance factors, $A_Z=1$.}
\tablenotetext{c}{N100 model \citep{2013MNRAS.429.1156S} with $A_{Fe}$=1.}
\tablenotetext{d}{N100 model with $A_{Fe}$=1, H He added at solar abundance.}
\tablenotetext{e}{N100 model with $A_{Fe}$=10.}
\tablenotetext{f}{N100 model with $A_{Fe}$=10, H He added at solar abundance.}
\end{deluxetable*}
\label{tab:tab1}

\clearpage

\begin{deluxetable*}{ccccccccccccccc}
\tabletypesize{\scriptsize}
\tablecaption{Abundances for the N100 model.} 
\tablewidth{0pt}
\tablehead{\\ 
\colhead{Z} & \colhead{Element} & \colhead{$M/M_{\odot}^{~a}$} &  \colhead{${N_{r,Fe}}^{b}$}   &   \colhead{${N/N_{sol}}^{c}$}   
&   \colhead{${A_{Z,Fe1}}^{d}$}  & \colhead{${A_{Z,Fe1,HHe}}^{e}$} & \colhead{${A_{Z,Fe10}}^{f}$}   & \colhead{${A_{Z,Fe10,HHe}}^{g}$}
}
\startdata
1&  H&7.33E-15&5.49E-13&5.49E-13&2.57E-17&1&2.57E-16&1 \\
2& He&8.31E-04&1.57E-02&1.60E-01&7.51E-06&1&7.51E-05&1 \\
3& Li&5.47E-19&5.95E-18&4.10E-07&1.92E-11&1.92E-11&1.92E-10&1.92E-10 \\
4&Be&4.44E-28&3.72E-27&2.64E-16&1.23E-20&1.23E-20&1.23E-19&1.23E-19 \\
5&  B&4.29E-12&3.00E-11&7.53E-02&3.52E-06&3.52E-06&3.52E-05&3.52E-05 \\
6&  C&3.04E-03&1.91E-02&5.26E+01&2.46E-03&2.46E-03&2.46E-02&2.46E-02 \\
7&  N&3.22E-06&1.73E-05&1.55E-01&7.25E-06&7.25E-06&7.25E-05&7.25E-05 \\
8&  O&1.01E-01&4.76E-01&5.60E+02&2.62E-02&2.62E-02&2.62E-01&2.62E-01 \\
9&  F&4.39E-11&1.74E-10&4.80E-03&2.25E-07&2.25E-07&2.25E-06&2.25E-06 \\
10& Ne&3.57E-03&1.34E-02&1.09E+02&5.08E-03&5.08E-03&5.08E-02&5.08E-02 \\
11& Na&3.74E-05&1.23E-04&5.74E+01&2.69E-03&2.69E-03&2.69E-02&2.69E-02 \\
12& Mg&1.54E-02&4.78E-02&1.26E+03&5.89E-02&5.89E-02&5.89E-01&5.89E-01 \\
13& Al&6.74E-04&1.89E-03&6.39E+02&2.99E-02&2.99E-02&2.99E-01&2.99E-01 \\
14& Si&2.87E-01&7.71E-01&2.17E+04&1.02E+00&1.02E+00&1.02E+01&1.02E+01 \\
15&  P&5.77E-04&1.41E-03&4.99E+03&2.33E-01&2.33E-01&2.33E+00&2.33E+00 \\
16&  S&1.15E-01&2.71E-01&1.67E+04&7.82E-01&7.82E-01&7.82E+00&7.82E+00 \\
17& Cl&2.13E-04&4.53E-04&1.44E+03&6.72E-02&6.72E-02&6.72E-01&6.72E-01 \\
18& Ar&1.96E-02&3.70E-02&1.02E+04&4.77E-01&4.77E-01&4.77E+00&4.77E+00 \\
19&  K&1.13E-04&2.18E-04&1.65E+03&7.73E-02&7.73E-02&7.73E-01&7.73E-01 \\
20& Ca&1.48E-02&2.79E-02&1.22E+04&5.70E-01&5.70E-01&5.70E+00&5.70E+00 \\
21& Sc&2.05E-07&3.44E-07&2.73E+02&1.28E-02&1.28E-02&1.28E-01&1.28E-01 \\
22& Ti&3.62E-04&5.70E-04&5.84E+03&2.73E-01&2.73E-01&2.73E+00&2.73E+00 \\
23&  V&1.05E-04&1.56E-04&1.56E+04&7.28E-01&7.28E-01&7.28E+00&7.28E+00 \\
24& Cr&1.03E-02&1.49E-02&3.19E+04&1.49E+00&1.49E+00&1.49E+01&1.49E+01 \\
25& Mn&1.33E-02&1.83E-02&7.46E+04&3.49E+00&3.49E+00&3.49E+01&3.49E+01 \\
26& Fe&7.40E-01&1.00E+00&2.14E+04&1.00E+00&1.00E+00&1.00E+01&1.00E+01 \\
27& Co&5.34E-04&6.84E-04&8.22E+03&3.85E-01&3.85E-01&3.85E+00&3.85E+00 \\
28& Ni&7.40E-02&9.52E-02&5.35E+04&2.50E+00&2.50E+00&2.50E+01&2.50E+01 \\
29& Cu&4.58E-07&5.44E-07&3.36E+01&1.57E-03&1.57E-03&1.57E-02&1.57E-02 \\
30& Zn&3.11E-06&3.59E-06&9.02E+01&4.22E-03&4.22E-03&4.22E-02&4.22E-02 \\
\enddata
\tablenotetext{a}{Masses from \cite{2013MNRAS.429.1156S}.}
\tablenotetext{b}{Number of nuclei relative to Fe. }
\tablenotetext{c}{Number of nuclei relative to solar abundance, taken to be `angr'.}
\tablenotetext{d}{Abundance factors $A_Z$  for the case of $A_{Z=26}$ (Fe) set to 1.}
\tablenotetext{e}{Same as case c, but with H and He added at solar abundance.}
\tablenotetext{f}{Abundance factors $A_Z$  for the case of $A_{Z=26}$ (Fe) set to 10.}
\tablenotetext{g}{Same as case c, but with H and He added at solar abundance.}
\end{deluxetable*}
\label{tab:n100}

\clearpage

\begin{deluxetable*}{cccccccccccc}
\tabletypesize{\scriptsize}
\tablecaption{Measured temperatures and $EM$ for type Ia SNRs$^{a}$} 
\tablewidth{0pt}
\tablehead{\\ 
\colhead{SNR ID} & \colhead {type} & \colhead {D(kpc)} &  \colhead{$EM(10^{58}$cm$^{-3}$)}  &   \colhead{norm}   & \colhead{ kT(keV)}  &   \colhead{Age(yr)$^{b}$} &   \colhead{density(cm$^{-3}$)$^{b}$}  &   \colhead{$n_e~t$(cm$^{-3}$s)$^{b}$} 
}
\startdata
G53.6-2.2&  Ia&  7.8&  0.022&    3.02E-04&  3.9&  44600&  0.096&  5.41E+11 \\
G299.2-2.9&  Ia&  5&  0.029&      9.70E-04&  1.36&  8800&  0.022&  2.45E+10 \\
G306.3-0.9&  Ia&  20&  1.8&      3.76E-03&  1.51&  12800&  2.5&  4.04E+12 \\
G315.4-2.3&  Ia&  2.8&  0.046&     4.91E-03&  3.04&  11600&  0.247&  3.62E+11 \\
G352.7-0.1&  Ia?&  7.5&  0.11&      1.64E-03&  3.2&  7600&  1.66&  1.59E+12 \\
average:&  &  &  &   2.32E-03&  2.602&  &  &  1.31E+12  \\
\enddata
\tablenotetext{a}{Values are from Tables 1 and 2 of \cite{2020ApJS..248...16L}, with references for  $EM$, norm and kT given in Table 1.}
\tablenotetext{b}{Values for age and density are from Table 2 of \cite{2020ApJS..248...16L}, with $n_e~t$ calculated using post-shock density of 4$n_e$.}
\end{deluxetable*}
\label{tab:snrem}

\clearpage
\begin{deluxetable*}{cccccccccccc}
\tabletypesize{\scriptsize}
\tablecaption{Fits to simulated spectra with added H and He$^{a}$} 
\tablewidth{0pt}
\tablehead{\\ 
\colhead{Spectrum}  & \colhead {$A_H$, $A_{He}$}  &  \colhead{$\chi^2$}  &   \colhead{probability}&  & \colhead {$A_H$, $A_{He}$}  &  \colhead{$\chi^2$}  &   \colhead{probability}  
}
\startdata
LSN  &2.6E-16,7.5E-05 &    110.4&    0.167 &&1, 1 &    109.8&    0.177  \\
 &4,4&    111.2&    0.153 & &8,8&    114.4&    0.109 \\
 &16,16&    119.2&    0.063 & &24,24&    122.4&    0.042 \\
&32,32&    124.6&    0.031 & &40,40&    122.4&    0.025 \\
\hline
MSN  &2.6E-16,7.5E-05 &    90.6&    0.662 &&1, 1 &    105.6&    0.258  \\
 &1.4,1.4&    113.1&    0.127 & &1.8,1.8&    119.2&    0.063 \\
 &2.2,2.2&    124.7&    0.033 & &2.6,2.6&    128.5&    0.018 \\
 \hline
HSN  &2.6E-16,7.5E-05 &    93.2&    0.590 &&2.6E-16,1 &    106.8&    0.233 \\
 &1,7.5E-05&    142.4&    1.9E-03 & &1,1&    176.1&    1.6E-06 \\
 &0.1,0.1&    96.0&    0.509 & &0.2,0.2&    100.3&    0.390 \\
  &0.3,0.3&    106.7&    0.236 & &0.4,0.4&    114.4&    0.109 \\
   &0.5,0.5&    123.0&    0.038 \\
\enddata
\tablenotetext{a}{LSN has exposure 10ks, norm 2.32 $10^{-3}$; MSN has exposure 20ks, norm 6.96 $10^{-3}$; HSN has exposure 20ks, norm 2.32 $10^{-2}$. All have kT=2.6 keV, $n_e~t=1.31~10^{12}$ and N100 abundances from column 8 of Table~\ref{tab:n100}. H and He are added to the fit model, as listed.}
\end{deluxetable*}
\label{tab:sim}

\clearpage

\end{document}